\documentclass[aps,floatfix,showpacs,superscriptaddress,showkeys,preprintnumbers]{revtex4-1} 

\usepackage{amssymb,amsmath,amsfonts,latexsym,graphicx,epsfig,here}

\newcommand{\Jpsi}{J\!/\!\psi}
\newcommand{\babar}{\mbox{\sl B\hspace{-0.4em} {\small\sl%
A}\hspace{-0.37em} \sl B\hspace{-0.4em} {\small\sl%
A\hspace{-0.02em}R}}\;}%\xspace}
\newcommand{\beaa}{\begin{eqnarray*}} 
\newcommand{\enaa}{\end{eqnarray*}}

\newcommand{\bea}{\begin{eqnarray}}
\newcommand{\ena}{\end{eqnarray}} 

\newcommand{\be}{\begin{equation}}
\newcommand{\en}{\end{equation}}

\newcommand{\nn}{\nonumber\\}
\newcommand{\IP}{\mbox{I}\!\mbox{P}}

\begin{document}

%\date{\today}

\preprint{DSF-5-2011}
\preprint{MZ-TH/11-09}

\title{One--photon decay of the tetraquark state 
$X(3872)\to \gamma +\Jpsi $ in a 
relativistic constituent quark model   with infrared confinement}

\author{Stanislav Dubnicka}
\affiliation{Institute of Physics
Slovak Academy of Sciences
Dubravska cesta 9
SK-845 11 Bratislava, Slovak Republic}

\author{Anna Z.~Dubnickova}
\affiliation{Comenius University
Department of Theoretical Physics
Mlynska Dolina
SK-84848 Bratislava, Slovak Republic }

\author{Mikhail A.~Ivanov}
\affiliation{Bogoliubov Laboratory of Theoretical Physics, 
Joint Institute for Nuclear Research, 141980 Dubna, Russia}

\author{J\"{u}rgen G.~K\"{o}rner}
\affiliation{Institut f\"{u}r Physik, Johannes Gutenberg-Universit\"{a}t,
D--55099 Mainz, Germany}

\author{Pietro Santorelli}
\affiliation{Dipartimento di Scienze Fisiche, Universit\`a di Napoli
Federico II, Complesso Universitario di Monte S. Angelo,
Via Cintia, Edificio 6, 80126 Napoli, Italy, and
Istituto Nazionale di Fisica Nucleare, Sezione di Napoli
}

\author{Gozyal G.~Saidullaeva}
\affiliation{Al-Farabi Kazak National University, 480012 Almaty, Kazakhstan}

\begin{abstract} 
We further explore the consequences of treating the X(3872) meson as a 
tetraquark bound state by analyzing its one-photon decay $X\to \gamma + \Jpsi $
in the framework of our approach developed in previous papers which 
incorporates quark confinement in an effective way. To introduce
electromagnetism we gauge a nonlocal effective Lagrangian describing the 
interaction of the X(3872) meson with its four constituent quarks by using the 
P-exponential path-independent formalism. 
We calculate the matrix element of the transition
$X\to \gamma+ \Jpsi $ and prove its gauge invariance.
We evaluate the $X\to \gamma + \Jpsi $ decay width and the 
longitudinal/transverse composition of the $\Jpsi$ in this decay. 
For a reasonable value of the size parameter of
the X(3872) meson we find consistency with the available
experimental data. We also calculate the helicity and multipole amplitudes
of the process, and describe how they can be obtained from the covariant
transition amplitude by covariant projection.
\end{abstract}


\maketitle

\section{Introduction}
\label{sec:intro}

This paper is a direct continuation of our previous work
\cite{Dubnicka:2010kz} where we have analyzed the strong decays of 
the charmonium--like state $X(3872)$ 
in the framework of our relativistic constituent quark model
which includes infrared confinement in an effective way \cite{Branz:2009cd}. 
In our approach the X(3872) meson is interpreted as a tetraquark state with 
the quantum numbers $J^{PC}=1^{++}$ as in \cite{Maiani:2004vq}. In this paper
we analyze the one-photon decay $X\to \gamma +\Jpsi$ in the same tetraquark
picture. The electromagnetic interaction is incorporated
into our relativistic nonlocal effective Lagrangian in a gauge invariant
way using the P-exponential path-independent formalism. 

We begin by collecting the experimental data relevant for our purposes. 
A narrow charmonium--like state $X(3872)$ was observed in 2003 
in the exclusive decay process $B^\pm\to K^\pm\pi^+\pi^-\Jpsi$
\cite{Choi:2003ue}. 
The  $X(3872)$ decays into $\pi^+\pi^-\Jpsi$ and has a mass of 
$m_X=3872.0 \pm 0.6 ({\rm stat}) \pm 0.5 ({\rm syst}) $
very close to the $m_{D^0}+m_{D^{\ast\,0}}=3871.81 \pm 0.25$ mass threshold 
\cite{PDG}.
Its width was found to be less than 2.3 MeV at $90\%$ confidence level.
The state was confirmed in B-decays by the \babar experiment 
\cite{Aubert:2004fc} 
and in $p\overline{p}$ production
by the Tevatron experiments CDF \cite{Acosta:2003zx} and D\O{}
\cite{Abazov:2004kp}. The most precise measurement up to now was done
in \cite{CDF} with $m_X=3871.61 \pm 0.16 \pm 0.19 $. The new average
mass given in \cite{Acosta:2003zx} is

\be
m_X=3871.51 \pm 0.22 \, {\rm MeV}.
\en  

The Belle Collaboration  has reported \cite{Abe:2005ix}
evidence for the decay modes $X(3872)\to\gamma +\Jpsi$ and to
$X\to\pi^+\pi^-\pi^0\Jpsi$:

\bea
{\cal B}(B\to X K)\cdot {\cal B}(X\to\gamma +\Jpsi)
 &=& 
(1.8 \pm 0.6\, {\rm (stat)} \pm 0.1\, {\rm (syst)}) \times 10^{-6}\,,
\nn
\frac{\Gamma(X\to\gamma + \Jpsi)}{\Gamma(X\to\pi^+\pi^-\Jpsi)} 
&=& 0.14\pm 0.05\,,
\nn
\frac{{\cal B}(X\to\pi^+\pi^-\pi^0\Jpsi)}{{\cal B}(X\to\pi^+\pi^-\Jpsi)}
&=& 1.0\pm 0.4\, {\rm (stat)}\pm 0.3\, {\rm (syst)}\,.
\label{eq:BELLE}
\ena

These observations imply strong isospin violation because the three-pion decay
proceeds via an intermediate $\omega$ meson with isospin 0 whereas
the two-pion decay proceeds via the intermediate $\rho$ meson with isospin 1.
It is evident that the two-pion decay via the intermediate $\rho$ meson is 
very difficult
to explain by using an interpretation of the $X(3872)$ as a
simple $c\bar c$ charmonium state with isospin 0. 

 In an analysis of $B^+\rightarrow \Jpsi\,\gamma\, K^+$ decays, the \babar
Collaboration  \cite{Aubert:2006aj} found evidence
for the radiative decay $X(3872)\rightarrow \gamma + \Jpsi$ 
with a statistical significance of $3.4\sigma$. 
They reported the following values for   the product of branching fractions 

\be
\mathcal{B}(B^+\rightarrow X K^+)\cdot
\mathcal{B}(X\rightarrow \gamma + \Jpsi) =
(3.3\pm 1.0\, {\rm (stat)}\pm 0.3 \, {\rm (syst)}) \times 10^{-6}\,.
\label{eq:BABAR_1}
\en

The Belle Collaboration reported \cite{Gokhroo:2006bt}
the first observation of a near-threshold enhancement 
in the  $D^0\bar D^0 \pi^0$ system from $B\rightarrow D^0\bar D^0 \pi^0 K$.
The enhancement peaks at a mass of 
$M=3875.2\pm  0.7_{-1.6}^{+0.3}\pm 0.8$~MeV. The branching fraction 
for events in the peak is 

\be
\mathcal{B}(B\rightarrow D^0\bar D^0 \pi^0 K)
=
(1.22\pm 0.31^{+0.23}_{-0.30}) \times 10^{-4}\,.
\label{eq:BELLE-DDpi}
\en

All available experimental data up to 2007 were analyzed in 
\cite{Klempt:2007cp}.
The authors found that \cite{Klempt:2007cp}

\bea
\mathcal{B}(B^+\rightarrow X K^+) &=& 1.30^{+0.20}_{-0.34}\times 10^{-4}\,,
\nn
\frac{\Gamma(X\to\gamma + \Jpsi)}{\Gamma(X\to\pi^+\pi^-\Jpsi)} 
&=& 0.22\pm 0.06\,.
\label{eq:Zaitsev}
\ena

 The \babar Collaboration found evidence for the decays 
$X\rightarrow \gamma +\Jpsi$ and $X\rightarrow \gamma + \psi(2S)$ in their data 
sample of the $B\rightarrow c\bar c\,\gamma K$ decays.
The measured products of branching fractions are \cite{:2008rn}

\bea 
{\cal B}(B^\pm\rightarrow X K^\pm)\cdot
{\cal B}(X\rightarrow \gamma + \Jpsi) 
&=& (2.8\pm 0.8\, {\rm (stat)}\pm 0.1\, {\rm (syst)})\times 10^{-6}\,,
\nn 
{\cal B}(B^\pm\rightarrow X K^\pm)\cdot
{\cal B}(X\rightarrow \gamma + \psi(2S)) 
&=& (9.5\pm 2.7\, {(\rm stat)}\pm 0.6\, {(\rm syst)})\times 10^{-6}.
\label{eq:BABAR-new}
\ena

There have been many theoretical attempts to unravel the structure of the 
$X(3872)$ and its decays. Many of
the theoretical predictions for the decay $X(3872)\to \gamma + \Jpsi$ published 
up to now are very model dependent. We mention some of them in turn.

All possible $1D$ and $2P$ $c\bar c$ assignments for the X(3872) were 
considered in \cite{Barnes:2003vb}. The authors obtained
$E1$ radiative widths for decays into charmonium $c\bar c$ states
as well as for some strong decays taking the experimental mass
as input. The conclusion was that many of the possible $J^{PC}$ assignments 
can be eliminated 
due to the smallness of the observed total width. The suggestion was that  
radiative transitions could be used to test the remaining $J^{PC}$ assignments.

Some tests of the hypothesis that the $X(3872)$ is a weakly bound  
$D^0\bar D^{0*}$ molecule state were suggested in \cite{Swanson:2004pp}.
It was proposed that measuring the $3\pi \Jpsi$, $\gamma + \Jpsi$, 
$\gamma + \psi'$, $\bar K K^*$, and $\pi\rho$
decay modes of the $X$  will serve as a definitive diagnostic tool to confirm or
to rule out the molecule hypothesis.

 Assuming that the $X(3872)$ state has the structure 
$(D^0\bar D^{0\,\ast}  -D^{0\,\ast}\bar D^{0})/\sqrt{2}$ with quantum
numbers $J^{PC} = 1^{++}$, 
the $X(3872)\to \gamma + \Jpsi$ decay width was calculated
using a phenomenological  Lagrangian approach \cite{Dong:2008gb}. 
The calculated value of the radiative decay width varied
from 125 KeV to 250 KeV depending on the model parameters.

QCD  sum rules were used in \cite{Nielsen:2010ij} to calculate the width  of the radiative decay  
of the meson  $X(3872)$, which was assumed to be a  mixture between charmonium  and  
exotic  molecular    $[c\bar{q}][q\bar{c}]$    states   with $J^{PC}=1^{++}$. 
In a  small range for  the values of the  mixing  angle, one obtains 

\be 
\frac{\Gamma(X\to \gamma +\Jpsi)}{\Gamma(X\to \Jpsi~\pi^+\pi^-)}=
0.19 \pm 0.13\,.
\en 
Our paper is organized as follows. In Sec.~II we gauge 
a nonlocal effective Lagrangian describing the interaction
of the X(3872) meson with its constituent quarks by using the 
P-exponential path-independent formalism developed in
\cite{Ivanov:1996pz,gauging}. 
In Sec.~III we calculate  the matrix element of the radiative transition
$X\to \gamma+ \Jpsi $ and prove its gauge invariance analytically.
In Sec.~IV we present the results of our numerical analysis.
First, we check numerically that the final amplitude is gauge invariant.
Second, we introduce infrared confinement as was done
in our previous papers Refs.~\cite{Dubnicka:2010kz,Branz:2009cd}
and evaluate the $X\to \gamma + \Jpsi $ decay width. 
Finally, in Sec.~V we summarize our results. In an Appendix we describe how
the two helicity or the two multipole amplitudes of the process can be 
obtained from the gauge invariant transition amplitude by covariant 
projection.

\section{Theoretical framework}

The effective interaction Lagrangians describing the coupling
of the charmonium-like meson such as the $X(3872)$ to four quarks,
and the coupling of the charmonium $\Jpsi$ state 
to its two constituent quarks are written in the form 
(see Ref.~\cite{Dubnicka:2010kz})

\bea
{\cal L}_{\rm int} &=&  g_X\,X_{q\,\mu}(x)\cdot J^\mu_{X_q}(x)
                    + g_{\Jpsi}\,\Jpsi_{\mu}(x)\cdot J^\mu_{\Jpsi}(x)  
 \qquad (q=u,d).
\label{eq:lag}
\ena     

The nonlocal interpolating quark currents read 

%\begin{widetext}
\bea
J^\mu_{X_q}(x) &=& \int\! dx_1\ldots \int\! dx_4 
\,\delta\left(x-\sum\limits_{i=1}^4 w_i x_i\right) 
\Phi_X\Big(\sum\limits_{i<j} (x_i-x_j)^2 \Big)
\nn
&\times&
\tfrac{1}{\sqrt{2}}\, \varepsilon_{abc}\varepsilon_{dec} \,
\Big\{\, [q_a(x_4)C\gamma^5 c_b(x_1)][\bar q_d(x_3)\gamma^\mu C \bar c_e(x_2)]
        +(\gamma^5\leftrightarrow \gamma^\mu)
\,\Big\},\nn
&&\nn
w_1&=&w_2  =\frac{m_c}{2(m_q+m_c)}\equiv \frac{w_c}{2}, \qquad
w_3 = w_4  =\frac{m_q}{2(m_q+m_c)}\equiv \frac{w_q}{2}, \nn
&&\nn
J^\mu_{\Jpsi}(y) &=&  \int\! dy_1 \int\! dy_2\,
\delta\left(y-\frac12 (y_1+y_2)\right) 
\Phi_{\Jpsi}\Big((y_1-y_2)^2 \Big) 
\bar c_a(y_1) \gamma^\mu c_a(y_2).
\label{eq:currents}
\ena 
%\end{widetext}

The matrix $C=\gamma^0\gamma^2$ is related to the charge conjugation 
matrix: $C=C^\dagger=C^{-1}=-C^T$, $C\Gamma^TC^{-1}=\pm\Gamma$, ($"+"$ for
$\Gamma=S,P,A$ and $"-"$ for $\Gamma=V,T$). 
We follow \cite{Maiani:2004vq} and take the tetraquark state to be a linear
superposition of the $X_u$ and $X_d$ states  according to

\bea  
X_l\equiv X_{\rm low} &=&\hspace{0.2cm}  X_u\, \cos\theta +  X_d\, \sin\theta,\nn
X_h\equiv X_{\rm high} &=& - X_u\, \sin\theta +  X_d\, \cos\theta.
\label{eq:mixing}
\ena

The coupling constant $g_X$ in Eq.~(\ref{eq:lag}) will be determined from
the compositeness condition $Z_{H}=0$ (see 
e.g. Refs.~\cite{SWH,Efimov:1993ei}). 
The compositeness condition requires that the renormalization constant 
$Z_H$ of the  elementary meson $X$ is set to zero, i.e.

\be
\label{eq:Z=0}
Z_H = 1-\Pi_H^\prime(p^{2}_{H}=m^2_H)=0,
\en
where $\Pi_X(p^2)$ is the scalar part of the meson mass operator and the
prime stands for the derivative w.r.t. $p^{2}_{H}$.
For the spin one states $X(3872)$ and $\Jpsi$ the compositeness condition reads
\bea
\Pi^{\mu\nu}_V(p) &=& g^{\mu\nu} \Pi_V(p^2) + p^\mu p^\nu \Pi^{(1)}_V(p^2),\nn
 \Pi_V(p^2) &=& \frac{1}{3}\left(g_{\mu\nu}-\frac{p_\mu p_\nu}{p^2}\right)
                \Pi^{\mu\nu}_V(p).
\label{eq:mass}
\ena
The $X$ meson mass operator can be calculated from the self--energy three-loop 
sunrise--type diagram with four quark-antiquark propagators. The calculation 
is described in more detail in Ref.~\cite{Dubnicka:2010kz}.

As in the case of baryons composed of three quarks it is convenient to 
transform to Jacobi coordinates in the integrals of
Eq.~(\ref{eq:currents}). In the case of four quarks one has

\bea
x_1&=&x  +  \frac{2w_2+w_3+w_4}{2\sqrt{2}} \rho_1 
           -  \frac{w_3-w_4}{2\sqrt{2}} \rho_2 
           +  \frac{w_3+w_4}{2}\rho_3\equiv x + \sum_{j=1}^3 c_{1j}\rho_j, \nn
x_2&=&x  -  \frac{2w_1+w_3+w_4}{2\sqrt{2}} \rho_1 
           -  \frac{w_3-w_4}{2\sqrt{2}} \rho_2 
           +  \frac{w_3+w_4}{2}\rho_3\equiv x + \sum_{j=1}^3 c_{2j}\rho_j, \nn
x_3&=&x  -  \frac{w_1-w_2}{2\sqrt{2}} \rho_1 
           +  \frac{w_1+w_2+2w_4}{2\sqrt{2}} \rho_2 
           -  \frac{w_1+w_2}{2}\rho_3\equiv x + \sum_{j=1}^3 c_{3j}\rho_j, \nn
x_4&=&x  -  \frac{w_1-w_2}{2\sqrt{2}} \rho_1 
           -  \frac{w_1+w_2+2w_3}{2\sqrt{2}} \rho_2 
           -  \frac{w_1+w_2}{2}\rho_3\equiv x + \sum_{j=1}^3 c_{4j}\rho_j, 
\label{eq:Jacobi-4}
\ena 
where $x=\sum\limits_{i=1}^4 x_i w_i$ and 
$\sum\limits_{1\le i< j\le 4} (x_i-x_j)^2 =\sum\limits_{i=1}^3 \rho_i^2.$ 
The inverse transformation reads
\be
\rho_1 = \sqrt{2}\,(x_1-x_2), \quad 
\rho_2 = \sqrt{2}\,(x_3-x_4), \quad 
\rho_3 = x_1+x_2-x_3-x_4.
\nonumber
\en  
In the case of two quarks as e.g. in the $\Jpsi$ case one has
\be
y_1 = y + \frac12 \rho, \quad  y_2= y - \frac12 \rho.
\label{eq:Jacobi-2}
\en  
One then has
\bea
J^\mu_{X_q}(x) &=& 
\int\! d\vec\rho\,\Phi_X(\vec\rho^{\,2})\,J^\mu_{4q}(x_1,\ldots,x_4),
\nn
J^\mu_{4q}(x_1,\ldots,x_4) &=&
\tfrac{1}{\sqrt{2}}\, \varepsilon_{abc}\varepsilon_{dec} \,
\Big\{\, [q_a(x_4)C\gamma^5 c_b(x_1)][\bar q_d(x_3)\gamma^\mu C \bar c_e(x_2)]
        +(\gamma^5\leftrightarrow \gamma^\mu)
\,\Big\},
\nn
&&\nn
J^\mu_{\Jpsi}(y) &=&  \int\! d\rho\, \Phi_{\Jpsi}(\rho^{2} ) J^\mu_{2q}(y_1,y_2),
\qquad
 J^\mu_{2q}(y_1,y_2) = \bar c_a(y_1) \gamma^\mu c_a(y_2),
\label{eq:currents-Jacobi}
\ena 
where $d\vec\rho=d\rho_1d\rho_2d\rho_3$ and $
\vec\rho^{\, 2}=\rho_1^2+\rho_2^2+\rho_3^2$.
The Jacobian is absorbed into the coupling $g_X$.

The gauge invariant interaction of a bound quark state with the 
electromagnetic field has been described in some detail 
in Ref.~\cite{Ivanov:1996pz}. For comprehensive
purposes we recall some of the key points of the gauging process. Since the 
$X(3872)$ and $\Jpsi$ mesons are neutral mesons we will discuss the charged 
quarks only.
The free Lagrangian of quarks is gauged in the standard manner by using minimal
substitution:

\be
\label{eq:free-quark}
\partial^\mu q \to (\partial^\mu - ie_q A^\mu) q, 
\qquad 
\partial^\mu \bar q \to (\partial^\mu +ie_q A^\mu)\bar q,
\en
where $e_q$ is the  quark's charge 
($e_u= \tfrac23\, e$, $e_d=-\,\tfrac13\, e$, etc.). 
Minimal substitution gives us the 
first piece of the electromagnetic interaction Lagrangian  

\be
\label{eq:quark_em}
{\cal L}^{\rm em (1)}_{\rm int}(x) = 
\sum_q e_q\, A_\mu(x)\,J^\mu_q(x), \qquad 
 J^\mu_q(x) = \bar q(x) \gamma^\mu q(x).
\en 

In order to guarantee gauge invariance of the nonlocal strong interaction 
Lagrangian, one multiplies each quark field $q(x_i)$ in  the relevant
quark current $J^\mu(x)$ given by Eq.~(\ref{eq:currents-Jacobi})
by a gauge field exponential according to

\bea 
\label{eq:gauging}
 q(x_i) &\to& e^{-ie_q I(x_i,x,P)}\,  q(x_i),\qquad  
\bar q(x_i) \to e^{ie_q I(x_i,x,P)}\,  \bar q(x_i),
\nn
&&\nn
I(x_i,x,P) &=& \int\limits_x^{x_i} dz_\mu A^\mu(z). 
\ena 
where $P$ is the path taken from $x$ to $x_i$.
It is readily seen that the full Lagrangian Eq.~(\ref{eq:lag}) is invariant 
under the local gauge transformations
\bea 
q(x_i) &\to& e^{ie_q f(x_i)} q(x_i),    \qquad
\bar q(x_i) \to e^{-ie_q f(x_i)} \bar q(x_i),
\nn  
A^\mu(z)&\to& A^\mu(z)+\partial^\mu f(z), \quad\text{so that} \quad
I(x_i,x,P)\to I(x_i,x,P) + f(x_i) - f(x).
\label{eq:gauge-group}
\ena

The second term of the electromagnetic interaction Lagrangian
${\cal L}^{em}_{\rm int; 2}$ arises when one expands the gauge exponential
in powers of $A_\mu$ up to the order of perturbation 
theory that one is considering. 
Superficially the results appear to depend on the path $P$
which connects the endpoints in the path integral in Eq~(\ref{eq:gauging}).
However, one needs to know only derivatives of the path
integrals when doing the perturbative expansion.
One can make use of the formalism developed in~\cite{gauging}
which is based on the path-independent definition of the derivative of 
$I(x,y,P)$: 
\begin{eqnarray}\label{eq:path1}
\lim\limits_{dx^\mu \to 0} dx^\mu 
\frac{\partial}{\partial x^\mu} I(x,y,P) \, = \, 
\lim\limits_{dx^\mu \to 0} [ I(x + dx,y,P^\prime) - I(x,y,P) ]\,,
\end{eqnarray}
where the path $P^\prime$ is obtained from $P$ by shifting the endpoint $x$
by $dx$.
Use of the definition (\ref{eq:path1}) 
leads to the key rule
\begin{eqnarray}\label{eq:path2}
\frac{\partial}{\partial x^\mu} I(x,y,P) = A_\mu(x)
\end{eqnarray}
which states that the derivative of the path integral $I(x,y,P)$ does 
not depend on the path $P$ originally used in the definition. The nonminimal 
substitution (\ref{eq:gauging}) is therefore completely equivalent to the 
minimal prescription as is evident from the identities (\ref{eq:path1}) or 
(\ref{eq:path2}). 
The method of deriving Feynman rules for the nonlocal coupling 
of hadrons to photons and quarks was worked out before in 
Refs.~\cite{Ivanov:1996pz,gauging} 
and will be discussed in the next section 
where we apply the formalism to the physical processes considered in this
paper. 

Expanding the Lagrangian  up to the first
order in $A^{\mu}$ one obtains 
\bea
{\cal L}^{\rm em (2)}_{\rm int}(x) &=& 
g_X\,X_{q\,\mu}(x)\cdot J^\mu_{X_q -\rm em}(x)
                    + g_{\Jpsi}\,\Jpsi_{\mu}(x)\cdot J^\mu_{\Jpsi-\rm em}(x)  
 \qquad (q=u,d),
\nn
&&\nn
J^\mu_{X_q -\rm em} &=& \int\! d\vec\rho\,\Phi_X(\vec\rho^{\,2})\,
J^\mu_{4q}(x_1,\ldots,x_4)\,
\Big\{ ie_q\,[I^{x_3}_x-I^{x_4}_x] +  ie_c\,[I^{x_2}_x-I^{x_1}_x] \Big\},
\nn
&&\nn
J^\mu_{\Jpsi -\rm em} &=& \int\! d\rho\,\Phi_{\Jpsi}(\rho^{\,2})\,J^\mu_{2q}(x_1,x_2)\,
ie_c\,[I^{x_1}_x-I^{x_2}_x],
\qquad
I^{x_i}_x  \equiv I(x_i,x,P). 
\label{eq:hadron_quark_em}
\ena 
 
In order to use the key rule Eq.~(\ref{eq:path2})
we take the Fourier-transforms for the vertex functions~$\Phi$
and quark fields~$q$ 
\bea
\Phi_X(\vec\rho^{\,2}) &=& \int\frac{d^4\vec\omega}{(2\pi)^4} 
\widetilde\Phi_X(-\vec\omega^{\,2})
e^{-i\vec\rho\vec\omega} = \widetilde\Phi_X(\vec\partial_\rho^{\,2})\,
\delta^{(4)}(\vec\rho),
\nn
\Phi_{\Jpsi}(\rho^{\,2}) &=& \int\frac{d^4\omega}{(2\pi)^4} 
\widetilde\Phi_{\Jpsi}(-\omega^{\,2})
e^{-i\rho\omega} = \widetilde\Phi_{\Jpsi}(\partial_\rho^{\,2})\,\delta^{(4)}(\rho),
\nn
 q(x_i) &=&  \int\frac{d^4p_i}{(2\pi)^4} e^{-ip_ix_i}\tilde q(p_i), \qquad
 \bar q(x_i) =  \int\frac{d^4p_i}{(2\pi)^4} e^{ip_ix_i} \tilde{\bar q}(p_i)\,.
\label{eq:Fourier}
\ena

One then writes down

\bea
J^\mu_{X_q -\rm em} &=& \prod\limits_{i=1}^4\int \frac{d^4p_i}{(2\pi)^4}
\widetilde J^\mu_{4q}(p_1,\ldots,p_4)
\!\int\! d\vec\rho\,\delta^{(4)}(\vec\rho)
\widetilde\Phi_X(\vec\partial_\rho^{\,2})\,
e^{-i(p_1x_1-p_2x_2-p_3x_3+p_4x_4)}
\Big\{ ie_q\,[I^{x_3}_x-I^{x_4}_x] +  ie_c\,[I^{x_2}_x-I^{x_1}_x] \Big\}
\nn
&=& \prod\limits_{i=1}^4\int \frac{d^4p_i}{(2\pi)^4}
\widetilde J^\mu_{4q}(p_1,\ldots,p_4)e^{-i(p_1-p_2-p_3+p_4)x}
\!\int\! d\vec\rho\,\delta^{(4)}(\vec\rho)
e^{-i\vec\rho\vec\omega}\widetilde\Phi_X(\vec D_\rho^{\,2})\,
\Big\{ ie_q\,[I^{x_3}_x-I^{x_4}_x] +  ie_c\,[I^{x_2}_x-I^{x_1}_x] \Big\},
\nn
&&\nn
J^\mu_{\Jpsi -\rm em} &=& \prod\limits_{i=1}^2\int \frac{d^4p_i}{(2\pi)^4}  
\widetilde J^\mu_{2q}(p_1,p_2)
\!\int\! d\rho\,\delta^{(4)}(\rho)\widetilde\Phi_{\Jpsi}(\partial_\rho^2)\,
e^{i(p_1x_1-p_2x_2)}
ie_c\,[I^{x_1}_x-I^{x_2}_x]
\nn
&=&\prod\limits_{i=1}^2\int \frac{d^4p_i}{(2\pi)^4}
\widetilde J^\mu_{2q}(p_1,p_2)e^{i(p_1-p_2)x}
\!\int\! d\rho\,\delta^{(4)}(\rho)
e^{ip\rho}\widetilde\Phi_{\Jpsi}(D_\rho^2)\,
ie_c\,[I^{x_1}_x-I^{x_2}_x],
\nn
&&\nn
D^\mu_{\rho_i} &=& \partial^\mu_{\rho_i}-i\omega^\mu_i, \qquad
D^\mu_{\rho}   = \partial^\mu_{\rho}+ip^\mu,
\label{eq:hadron_quark_em_1}
\ena 
where

\bea
\omega_1 &=& c_{11}p_1-c_{21}p_2-c_{31}p_3+c_{41}p_4,
\nn
\omega_2 &=& c_{12}p_1-c_{22}p_2-c_{32}p_3+c_{42}p_4,
\nn
\omega_3 &=& c_{13}p_1-c_{23}p_2-c_{33}p_3+c_{43}p_4,
\nn
p &=& \tfrac12 (p_1+p_2).
\label{eq:omega}
\ena

Finally, we employ a convenient identity which was proven in 
\cite{Ivanov:1996pz}. The identity reads 

\bea
F(D^2_{\rho_j}) I_x^{x_i} &=&
\int\limits_0^1\! d\tau F^\prime( \tau D^2_{\rho_j} - (1-\tau)\omega^2_j )\,
c_{ij}\,\Big( \partial^\nu_{\rho_j}A_\nu(x_i) - 2\,i\,\omega^\nu_j A_\nu(x_i) \Big)
+F(-\omega_j^2)I^{x_i}_x 
\label{eq:key-formula}\,.
\ena
The identity holds for any function $F(z)$ that is analytical at $z=0$.

One obtains

\bea
J^\mu_{X_q -\rm em}(x) &=& \prod\limits_{i=1}^4 \!\int\!d^4x_i\!\int\! d^4y \,
J^\mu_{4q}(x_1,\ldots,x_4)\,A_\rho(y)\,E_X^\rho(x;x_1,\ldots,x_4,y),
\label{eq:X-em-2}\\
E_X^\rho(x;x_1,\ldots,x_4,y) &=&  
\prod\limits_{i=1}^4\!\int\!\frac{d^4p_i}{(2\pi)^4}\!\int\!\frac{d^4r}{(2\pi)^4}
e^{-ip_1(x-x_1)+ip_2(x-x_2)+ip_3(x-x_3)-ip_4(x-x_4)-ir(x-y)}\,
\widetilde E_X^\rho(p_1,\ldots,p_4,r),
\nn
\widetilde E_X^\rho(p_1,\ldots,p_4,r) &=& \int\limits_0^1 d\tau
\sum\limits_{j=1}^3
\Big\{
e_c\left[
-\widetilde\Phi_X^\prime(-z_{1j})\,l^\rho_{1j}
+\widetilde\Phi_X^\prime(-z_{2j})\,l^\rho_{2j}
\right]
+
e_q\left[
-\widetilde\Phi_X^\prime(-z_{4j})\,l^\rho_{4j}
+\widetilde\Phi_X^\prime(-z_{3j})\,l^\rho_{3j}
\right]
\Big\}
\nn
&&\nn
l_{ij} &=& c_{ij}\,(c_{ij}r + 2\,\omega_j), \qquad (i=1,\ldots,4;j=1,\,\ldots,3),
\nn
z_{i1} &=& \tau\,(c_{i1}r+\omega_1)^2+(1-\tau)\,\omega_1^2+\omega_2^2+\omega_3^2,
\nn
z_{i2} &=& (c_{i1}r+\omega_1)^2 + \tau\,(c_{i2}r+\omega_2)^2
         + (1-\tau)\,\omega_2^2 + \omega_3^2,
\nn
z_{i3} &=& (c_{i1}r+\omega_1)^2 + (c_{i2}r+\omega_2)^2 + \tau\,(c_{i3}r+\omega_3)^2
         +(1-\tau)\,\omega_3^2.
\nonumber
\ena

\bea
J^\nu_{\Jpsi-\rm em}(y) &=&
\int\!d^4y_1\!\int\!d^4y_2\!\int\!d^4z\, J^\nu_{2q}(y_1,y_2)
\,A_\rho(z)\,E_{\Jpsi}^\rho(y;y_1,y_2,z) \,,
\label{eq:X-em-3}\\
E_{\Jpsi}^\rho(y;y_1,y_2,z) &=&  
\int\!\frac{d^4p_1}{(2\pi)^4}\!\int\!\frac{d^4p_2}{(2\pi)^4}
\!\int\!\frac{d^4q}{(2\pi)^4}
e^{-ip_1(y_1-y)+ip_2(y_2-y)+iq(z-y)}\,
\widetilde E_{\Jpsi}^\rho(p_1,p_2,q)\,,
\nn
\widetilde E_{\Jpsi}^\rho(p_1,p_2,q) &=& e_c\int\limits_0^1 d\tau\,
\Big\{
-\widetilde\Phi_{\Jpsi}^\prime(-z_{-})\,l^\rho_{-}
-\widetilde\Phi_{\Jpsi}^\prime(-z_{+})\,l^\rho_{+}
\Big\}\,,
\nn
z_\mp &=& \tau\,(p\mp \tfrac12 q)-(1-\tau)\,p^2\,, \qquad
l_{\mp} = p \mp \tfrac14 q\,, \qquad p = \tfrac12\, (p_1+p_2)\,.
\nonumber
\ena 

For calculational convenience we will choose a simple Gaussian form for the 
vertex function $\bar \Phi_X(-\,\Omega^2)$. 
The minus sign in the argument of the Gaussian function is chosen to emphasize 
that we are working in Minkowski space. One has
\be
\bar \Phi_X(-\,\Omega^2) 
= \exp\left(\Omega^2/\Lambda_X^2\right)
\label{eq:Gauss}
\en
where the parameter $\Lambda_X$ characterizes the size of the X meson.
Since $\Omega^2$ turns into 
$-\,\Omega^2$ in Euclidean space the form
(\ref{eq:Gauss}) has the appropriate fall-off behavior in the Euclidean region.
We emphasize that any choice for  $\Phi_X$ is appropriate
as long as it falls off sufficiently fast in the ultraviolet region of
Euclidean space to render the corresponding Feynman diagrams ultraviolet 
finite. As mentioned before we shall choose a Gaussian form for $\Phi_X$
in our numerical calculation for calculational convenience.

\section{Matrix element for the decay \boldmath{$X\to \gamma +\Jpsi $} }

\begin{figure}[ht]
\begin{tabular}{lr}
\includegraphics[scale=0.4]{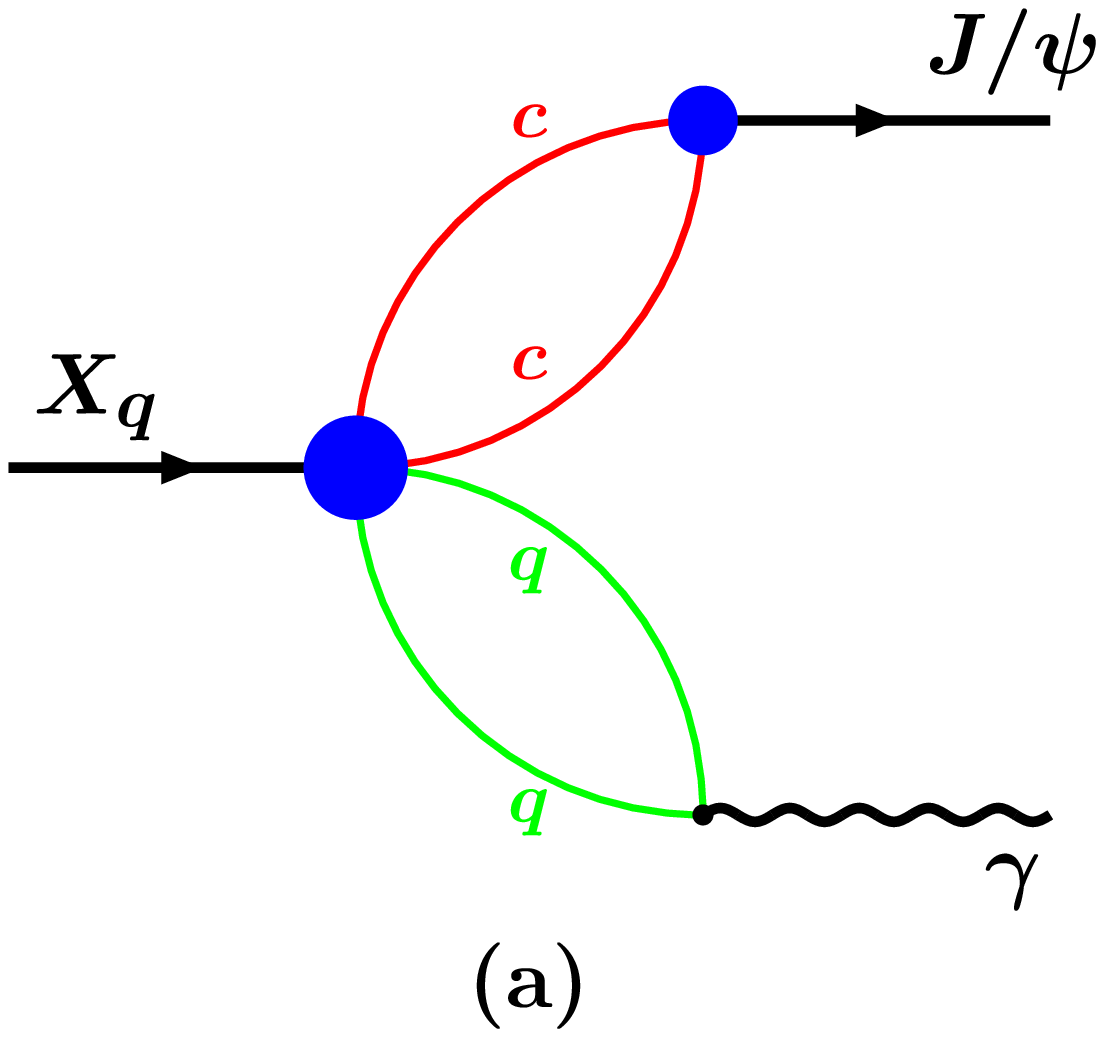} & 
\includegraphics[scale=0.4]{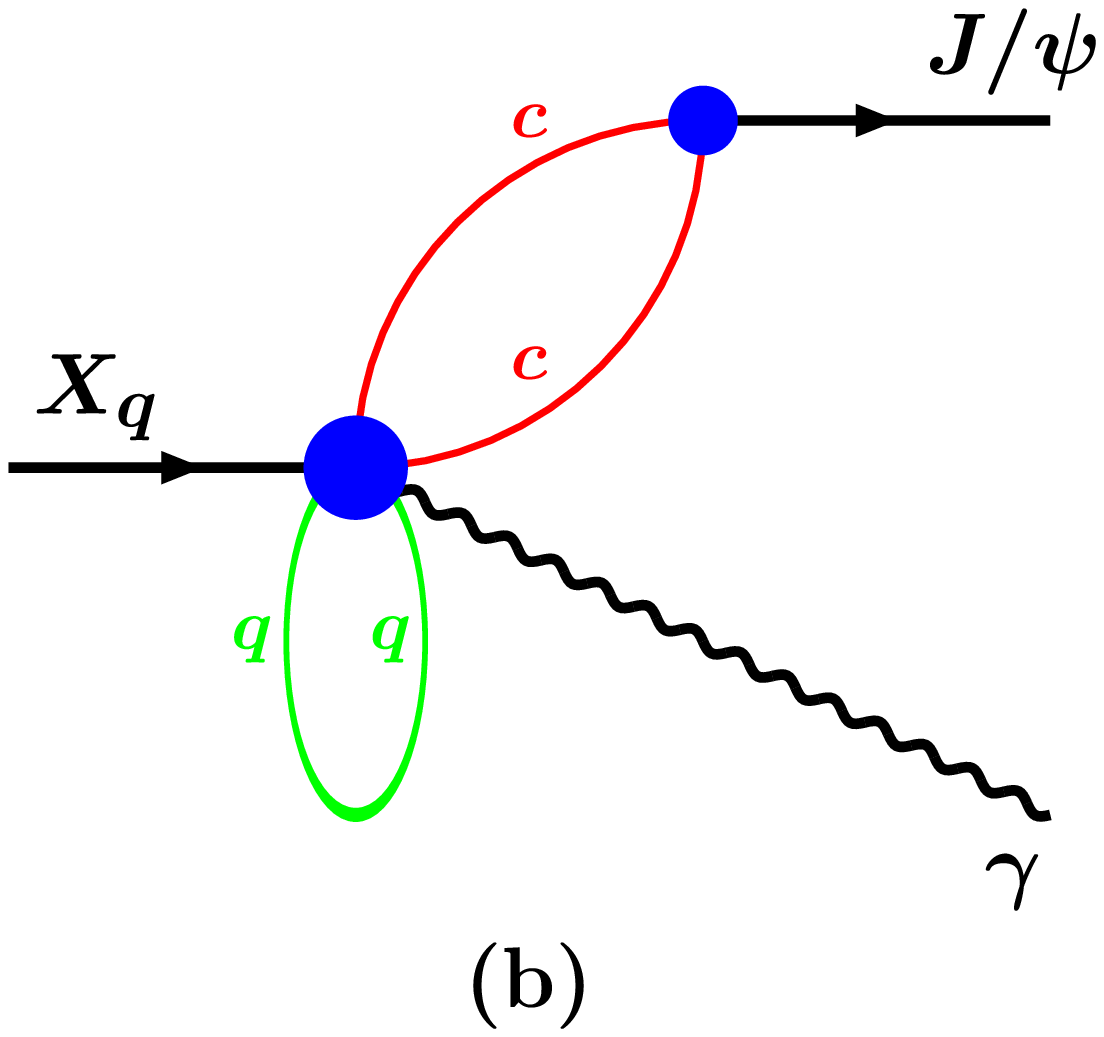}\\[2ex]
\includegraphics[scale=0.4]{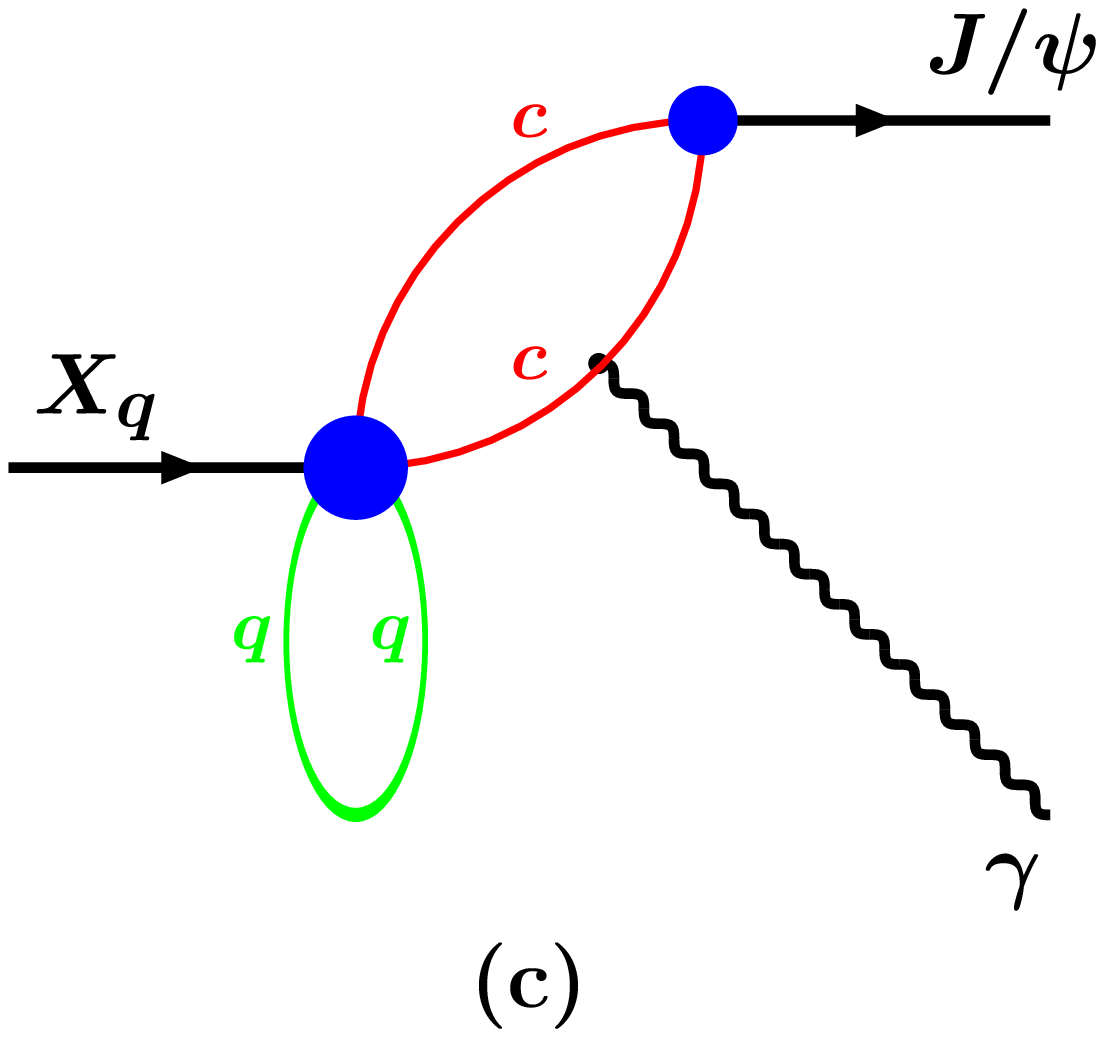} & 
\includegraphics[scale=0.4]{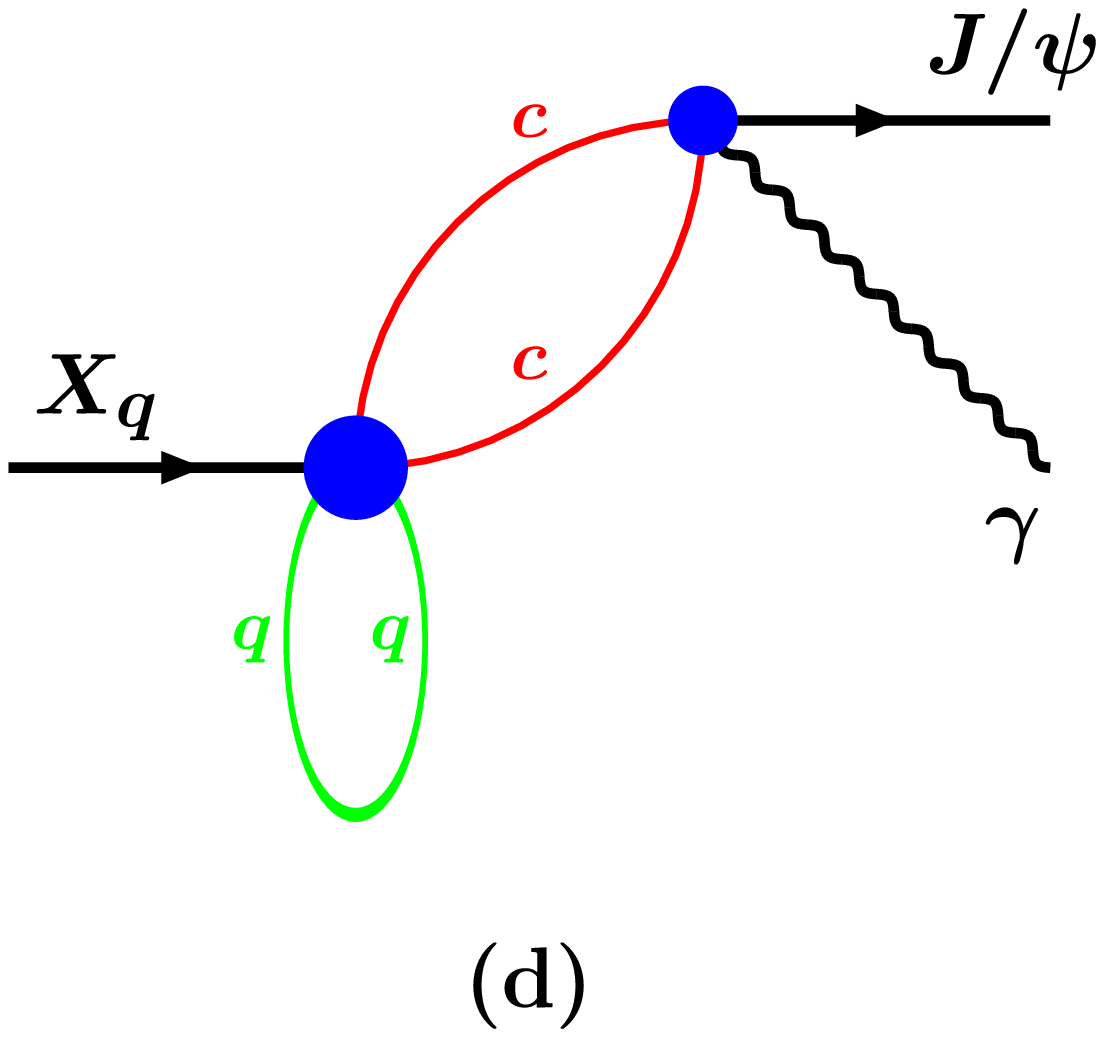}
\end{tabular}
\caption{
Feynman diagrams describing the decay
$X\to \gamma +  \Jpsi$.
}
\label{fig:decay}
\end{figure}

The matrix element of the decay
$X(3872)\to \gamma + \Jpsi$ can be calculated from 
the Feynman diagrams shown in Fig.~\ref{fig:decay}.
The invariant matrix element for the decay is given by

\be
M(X_q(p)\to\Jpsi(q_1)\,\gamma(q_2)) =
i(2\pi)^4\delta^{(4)}(p-q_1-q_2)\,
\varepsilon_X^\mu\,\varepsilon^\rho_\gamma\,\varepsilon^\nu_{\Jpsi}\,
T_{\mu\rho\nu}(q_1,q_2)\,,
\label{eq:matrix-element}
\en
where
\bea
T_{\mu\rho\nu}(q_1,q_2) &=&\sum\limits_{i=a,b,c,d}T_{\mu\rho\nu}^{(i)}(q_1,q_2)\,,
\nn
T^{(a)}_{\mu\rho\nu} &=&
6\,\sqrt{2}\,g_X\, g_{\Jpsi}\,e_q\,
\int\!\!\frac{d^4k_1}{(2\pi)^4i}\int\!\!\frac{d^4k_2}{(2\pi)^4i}
\widetilde\Phi_X\Big(-K_a^2\Big) 
\widetilde\Phi_{\Jpsi}\Big(-(k_1+\tfrac12 q_1)^2\Big)
\nn
&\times&
\tfrac12\, {\rm tr}\Big[\gamma_5 S_c(k_1)\gamma_\nu S_c(k_1+q_1)\gamma_\mu
S_q(k_2)\gamma_\rho S_q(k_2+q_2) - (\gamma_5\leftrightarrow \gamma_\mu)\Big]\,, 
\nn
K^2_a &=& \tfrac12\,(k_1+\tfrac12\,q_1)^2 +  \tfrac12\,(k_2+\tfrac12\,q_2)^2
          + \tfrac14\,(w_q q_1 - w_c q_2)^2\,,
\nn
&&\nn
T^{(b)}_{\mu\rho\nu} &=&
6\,\sqrt{2}\,g_X\, g_{\Jpsi}\,
\int\!\!\frac{d^4k_1}{(2\pi)^4i}\int\!\!\frac{d^4k_2}{(2\pi)^4i}
\widetilde\Phi_{\Jpsi}\Big(-(k_2+\tfrac12 q_1)^2\Big)
{\widetilde E}_{X\,\rho}(p_1,\ldots,p_4,r)
\nn
&\times&
\tfrac12\, {\rm tr}\Big[\gamma_5 S_q(k_1)\gamma_\mu S_c(k_2)\gamma_\nu
S_c(k_2+q_1) - (\gamma_5\leftrightarrow \gamma_\mu)\Big]\,, 
\nn
&&
p_1=k_2, \qquad p_2=k_2+q_1, \qquad p_3=p_4=-k_1, \qquad r=-q_2\,,
\nn
&&\nn
T^{(c)}_{\mu\rho\nu} &=&
6\,\sqrt{2}\,g_X\, g_{\Jpsi}\,e_c
\int\!\!\frac{d^4k_1}{(2\pi)^4i}\int\!\!\frac{d^4k_2}{(2\pi)^4i}
\widetilde\Phi_X\Big(-K_c^2\Big) 
\widetilde\Phi_{\Jpsi}\Big(-(k_2 + q_2 + \tfrac12 q_1)^2\Big)
\nn
&\times&
\tfrac12\, {\rm tr}\Big[\gamma_5 S_q(k_1)\gamma_\mu S_c(k_2)\gamma_\rho
S_c(k_2+q_2)\gamma_\nu S_c(k_2+p) 
- (\gamma_5\leftrightarrow \gamma_\mu)\Big]\,, 
\nn
K^2_c &=& \tfrac12\,k_1^2 +  \tfrac12\,(k_2+\tfrac12\,p)^2
          + \tfrac14\,w_q^2 p^2\,,
\nn
&&\nn
T^{(d)}_{\mu\rho\nu} &=&
6\,\sqrt{2}\,g_X\, g_{\Jpsi}\,e_c
\int\!\!\frac{d^4k_1}{(2\pi)^4i}\int\!\!\frac{d^4k_2}{(2\pi)^4i}
\widetilde\Phi_{X}\Big(-K_c^2\Big)
{\widetilde E}_{\Jpsi\,\rho}(p_1,p_2,q)
\nn
&\times&
\tfrac12\, {\rm tr}\Big[\gamma_\mu S_q(k_1)\gamma_5 S_c(k_2)\gamma_\nu
S_c(k_2+p) - (\gamma_5\leftrightarrow \gamma_\mu)\Big]\,, 
\nn
&&
p_1=-k_2-p, \qquad p_2=-k_2, \qquad q=-q_2\,.
\nonumber
\ena

We have analytically checked on the gauge invariance of the unintegrated 
transition matrix element by contraction with the photon momentum $q_{2}$
which yields $q^\rho_2  T_{\mu\rho\nu}(q_1,q_2)=0$ using the identities

\bea
&&
S(k_2)\!\not\! q_2 \,S(k_2+q_2) =  S(k_2+q_2) - S(k_2)\,,
\nn
&&\nn
&&
\int\limits_0^1 d\tau\,
\widetilde\Phi^\prime(-\tau\,a-(1-\tau)\,b)\,(a-b)=
\widetilde\Phi(-b)-\widetilde\Phi(-a).
\nonumber
\ena

\section{Numerical results}

The evaluation of the loop integrals in Eq.~(\ref{eq:matrix-element})
proceeds as described in our previous paper \cite{Dubnicka:2010kz}.
If one takes the on-mass shell conditions into account

\be
\varepsilon_X^\mu  p_\mu=0, \qquad \varepsilon_{\Jpsi}^\nu q_{1 \nu}=0, \qquad
\varepsilon_\gamma^\rho q_{2 \rho}=0
\label{eq:on-mass-shell}
\en
one can write down five seemingly independent Lorentz structures
 
\be
\label{basis}
T_{\mu\rho\nu}(q_1,q_2) = \varepsilon_{q_2\mu\nu\rho} (q_1\cdot q_2)\,W_1
                      +\varepsilon_{q_1q_2\nu\rho} q_{1 \mu}\,W_2
                      +\varepsilon_{q_1q_2\mu\rho} q_{2 \nu}\,W_3
                      +\varepsilon_{q_1q_2\mu\nu} q_{1 \rho}\,W_4
                      +\varepsilon_{q_1\mu\nu\rho} (q_1\cdot q_2)\,W_5\,.
\en

Using the gauge invariance condition 
\be
q_2^\rho T_{\mu\rho\nu}=(q_1\cdot q_2)\varepsilon_{q_1q_2\mu\nu}(W_{4}+W_{5})
=0
\en
one has $W_{4}=-W_{5}$ which reduces the set of independent covariants to four:
\be
\label{basis2}
T_{\mu\rho\nu}(q_1,q_2) = (q_1\cdot q_2)\,\varepsilon_{q_2\mu\nu\rho}\,W_1
                      +\varepsilon_{q_1q_2\nu\rho} q_{1 \mu}\,W_2
                      +\varepsilon_{q_1q_2\mu\rho} q_{2 \nu}\,W_3
 +\Big(
   \varepsilon_{q_1q_2\mu\nu} q_{1 \rho} 
- (q_1\cdot q_2)\varepsilon_{q_1\mu\nu\rho}\Big)\,W_4\,.
\en
The gauge invariance condition $W_{4}=-W_{5}$
provides for a numerical check on the gauge invariance of our 
calculation as described further on. 

However, there are two nontrivial relations among the four covariants 
which can be derived by noting \cite{Korner:2003zq} that the tensor
\be
T_{\mu[\nu_{1}\nu_{2}\nu_{3}\nu_{4}\nu_{5}]}=
g_{\mu\nu_{1}}\varepsilon_{\nu_{2}\nu_{3}\nu_{4}\nu_{5}}
+{\rm cycl.}(\nu_{1}\nu_{2}\nu_{3}\nu_{4}\nu_{5})
\en
vanishes in four dimensions since it is totally antisymmetric in the five 
indices $(\nu_{1},\nu_{2},\nu_{3},\nu_{4},\nu_{5})$. Upon contraction with
$q_{1}^{\mu}q_{1}^{\nu_{1}}q_{2}^{\nu_{2}}$ and 
$q_{2}^{\mu}q_{1}^{\nu_{1}}q_{2}^{\nu_{2}}$ one finds (between polarization
vectors)
\bea
&&
  q_1^2 \varepsilon_{q_2\mu\nu\rho}
+ \varepsilon_{q_1q_2\nu\rho} q_{1 \mu}
+\Big(\varepsilon_{q_1q_2\mu\nu} q_{1 \rho} - (q_1\cdot q_2) \varepsilon_{q_1\mu\nu\rho}\Big)
=0\,,
\label{constraint1}\\
&&
(q_1\cdot q_2)\varepsilon_{q_2\mu\nu\rho}
-\varepsilon_{q_1q_2\nu\rho} q_{1 \mu}
-\varepsilon_{q_1q_2\mu\rho} q_{2 \nu}
=0\,.
\label{constraint2}
\ena
The two conditions reduce the set of independent covariants to two.
This is the appropriate number of independent covariants since the photon
transition is described by two independent amplitudes as e.g. by the $E1$ and
$M2$ transition amplitudes.

Using the two constraint Eqs. (\ref{constraint1}) and (\ref{constraint2})
the expansion (\ref{basis2}) can be written in the form
\be
\label{basis3}
T_{\mu\rho\nu}=
\left(W_{1}+W_{3}-\frac{m_{\Jpsi}^{2}}{(q_{1}\cdot q_{2})}W_{4}\right) 
\,\varepsilon_{q_1q_2\mu\rho} q_{2 \nu}
+\left(W_{1}+W_{2}-\left (1+\frac{m_{\Jpsi}^{2}}{(q_{1}\cdot q_{2})}\right )
W_{4}\right) \,\varepsilon_{q_1q_2\nu\rho} q_{1 \mu}.
\en 
By comparing with the corresponding expressions in the Appendix one notes that 
the first and second terms in (\ref{basis3}) describe transitions into the 
longitudinal and transverse components of the $\Jpsi$.

The quantities $W_i$ are represented by the four-fold integrals
\be
W_i=\int\limits_0^\infty\! dt\! \int\limits_0^1\! d^3 \beta\, 
F_{i}(t,\beta_1,\beta_2,\beta_3)\,,
\label{eq:structures}
\en
where we have suppressed the additional dependence of the integrand $F_{i}$
on the set of variables $p^2,q_1^2,q_2^2;m_q,m_c,s_X,s_{\Jpsi}$ with
$s_X=1/\Lambda_X^2$ and $s_{\Jpsi}=1/\Lambda_{\Jpsi}^2$. 
The integrals in Eq.~(\ref{eq:structures})  have branch points
at $p^2=4(m_q+m_c)^2$ [diagram in Fig.~\ref{fig:decay}-a]
and at  $p^2=4 m_c^2$ [diagrams in Figs.~\ref{fig:decay}-b,c,d].
At these points the integrals become nonanalytical in the conventional sense
when $t\to \infty$. In order to check on the gauge invariance 
of the amplitude $T_{\mu\rho\nu}(q_1,q_2)$,
we have taken the X-meson momentum squared to be below the closest
unitarity threshold, i.e. $p^2<4m_c^2$. We have checked explicitly that, 
for $m_X=3.1$ GeV and $m_{\Jpsi}=2.9$ GeV, the gauge condition $W_4=-W_5$ 
is numerically satisfied to very high accuracy. Note that the gauge invariance condition is
independent of the overall couplings $g_X$ and $g_{\Jpsi}$ and thus the
numerical check can be done irrelevant of their values. 

In the next step we introduce an infrared cutoff $1/\lambda^2$ on the upper 
limit of the $t$-integration in Eq.~(\ref{eq:structures}). In this manner one 
removes all possible nonanalytic structures and thereby one obtains entire 
functions for the amplitudes, i.e. one has effectively instituted quark 
confinement, see Refs.~\cite{Branz:2009cd,Dubnicka:2010kz}. The value of 
$\lambda=181$~MeV was found by fitting the calculated basic quantities
to the experimental data. However, for such a value of $\lambda$ the 
contributions coming from the bubble diagrams in Figs.~\ref{fig:decay}-b,c,d 
blow up at $p^2=m^2_X$ 
compared with the contribution from the diagram Fig.~\ref{fig:decay}-a. 
The bubble diagrams are needed
only to guarantee the gauge invariance of the matrix element. For physical
applications one should take into account only the gauge invariant part of the
diagram Fig.~\ref{fig:decay}-a. 

It is convenient to present the decay width via helicity or multipole
amplitudes. The projection of the Lorentz amplitudes to
the helicity amplitudes is given in Appendix.
One has 
\be
\Gamma(X\to \gamma\,\Jpsi) =\frac{1}{12\pi}\,
 \frac{|\vec{q}_2|}{m_X^2}\,\Big( |H_L|^2 +|H_{T}|^{2}\Big)
= \frac{1}{12\pi}\,\frac{|\vec{q}_2|}{m_X^2}\,\Big( |A_{E1}|^2 +|A_{M2}|^{2}\Big)\,,
\label{eq:width}
\en
where the helicity amplitudes $H_L$ and $H_T$ are expressed in terms
of the Lorentz amplitudes as
\bea
H_L &=& i\frac{m_X^{2}}{m_{\Jpsi}}|\vec q_2|^{2}
     \Big[ W_1 + W_3-\frac{m^2_{\Jpsi}}{m_X|\vec q_2|}W_4\Big]\,,
\nn
H_T &=& -im_{X}|\vec q_2|^{2}
      \Big[W_1+W_2
        -\Big(1+\frac{m^2_{\Jpsi}}{m_X|\vec q_2|}\Big)\,W_4\Big]\,,
\nn
&& |\vec q_2|=\frac{m_X^2-m^2_{\Jpsi}}{2m_X}\,.
 \ena
The $E1$ and $M2$ multipole amplitudes are obtained via 
$A_{E1/M2}=(H_{L}\mp H_{T})/\sqrt{2}$\,. If we choose $\Lambda_{X}= 3.0$ GeV
for the size parameter of the $X(3872)$ we obtain $A_{M2}/A_{E1}=0.11$,
i.e. the electric multipole amplitude $A_{E1}$ dominates the transition, as 
expected. Nevertheless our predicted angular decay distribution 
$W(\vartheta)\sim 1-0.52\cos^{2}\vartheta$ differs noticeably from its form
$W(\vartheta)\sim 1-0.333\cos^{2}\vartheta$ for $E1$ dominance. It would be 
interesting to experimentally check on this prediction of our tetraquark 
model.

\begin{figure}[ht]
\includegraphics[scale=0.40]{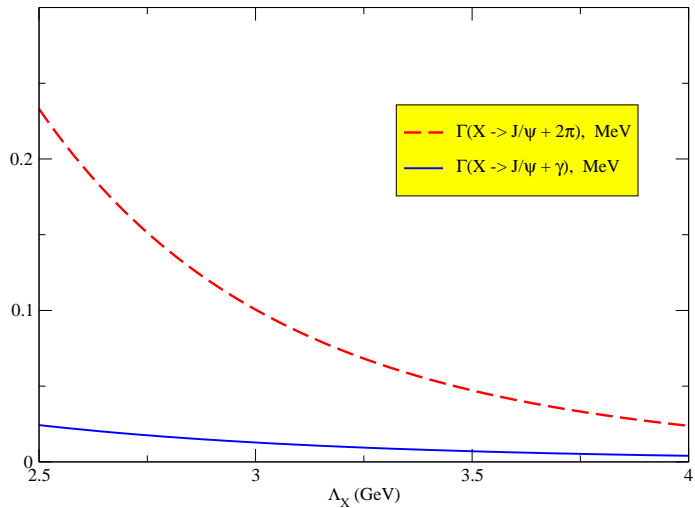}
\caption{The dependence of the decay widths 
$\Gamma(X_l\to \gamma + \Jpsi)$ 
and $\Gamma(X_l\to \Jpsi\, 2\pi)$ on the size parameter $\Lambda_X$.
}
\label{fig:width}
\end{figure}

In Fig.~\ref{fig:width} we show a plot of the size parameter dependence of
the decay width $\Gamma(X_l\to \Jpsi+\gamma)$ together with the decay width
$\Gamma(X_l\to \Jpsi+2\pi)$ taken from~\cite{Dubnicka:2010kz}.
We correct an error of Ref.~\cite{Dubnicka:2010kz}
in the normalization condition of the X meson, which led to
a $\lesssim 30\%$ underestimate of the strong decay widths.
Both decay widths become smaller as the size parameter increases. Note that the 
radiative decay width for $X_h=-X_u\sin\theta+X_d\cos\theta$ is almost an order 
of magnitude smaller than that for $X_l=X_u\cos\theta+X_d\sin\theta$. 
If one takes $\Lambda_X\in (3,4)$~GeV with the central value 
$\Lambda_X=3.5$~GeV then our prediction for the ratio of widths reads

\be
\frac{\Gamma(X_l\to \gamma+ \Jpsi)}
     {\Gamma(X_l\to \Jpsi+2\pi)}\Big|_{\rm theor} = 0.15\pm 0.03 
\label{eq:theory}
\en
which fits very well the experimental data from the Belle Collaboration
\cite{Abe:2005ix}
\be
\frac{\Gamma(X\to \gamma + \Jpsi)}
     {\Gamma(X\to \Jpsi\, 2\pi)}     = \left\{
 \begin{array}{rl}
0.14 \pm 0.05   & \mbox{ Belle\,\cite{Abe:2005ix} } \\[2ex]
0.22 \pm 0.06   & \text{ \babar\,\cite{Klempt:2007cp} } 
\end{array}\right.
\label{eq:expt}
\en

\vspace*{1cm}

\section{Summary and conclusion} 

We have used our relativistic constituent quark model
which includes infrared confinement in an effective way to calculate the
radiative decay $X\to \gamma + \Jpsi $. We take the X(3872) meson to be a 
tetraquark state with the quantum numbers $J^{PC}=1^{++}$. 
In order to introduce electromagnetic interactions we have gauged a nonlocal 
effective Lagrangian which describes the interaction
of the X(3872) meson with its four constituent quarks by using the 
P-exponential path-independent formalism. 
We have calculated  the matrix element of the transition
$X\to \gamma + \Jpsi $ and have shown its gauge invariance.
We have evaluated the $X\to \gamma + \Jpsi $ decay width and the polarization
of the $\Jpsi$ in the decay.
The calculated decay width is consistent with the available
experimental data for reasonable values of the size parameter of
the X(3872) meson.

\begin{acknowledgments}

This work was supported by the DFG grant KO 1069/13-1,
the Heisenberg-Landau program, the Slovak aimed project at JINR 
and the grant VEGA No.2/0009/10. M.A.I. also appreciates the partial 
support of the Russian Fund of Basic Research Grant No. 10-02-00368-a. 

\end{acknowledgments}

\appendix

\section{Helicity and multipole amplitudes}

The material presented in this Appendix is adapted from similar material
written down in \cite{Korner:1982vg} in a slightly different context. 
There are two independent helicity amplitudes
$H_{\lambda_{X};\lambda_{\gamma}\lambda_{\Jpsi}}$ which we denote by 
$H_{i}\,\,(i=L,T)$ according to the helicity of the final meson state $\Jpsi$, 
where $\lambda_{\Jpsi}=0$ and $\lambda_{\Jpsi}=\pm1$ stand for the longitudinal
and transverse helicities of the $\Jpsi$. From parity one has 
$H_{+;-0}=-H_{-;+0}=H_{L}$ and $H_{0;++}=-H_{0;--}=H_{T}$.

We seek a covariant representation for the longitudinal and transverse
projectors $\IP_{L,T}^{\mu \rho \nu}$ which, when applied to the transition 
amplitude $T_{\mu \rho \nu}$, project onto the helicity amplitudes
$H_{L,T}$ according to 
\be
H_{i}= \IP_{i}^{\mu \rho \nu}T_{\mu \rho \nu}, \qquad (i=L,T).
\label{eq:helicity}
\en
The projectors are defined by 
\bea
\IP_{L}^{\mu \rho \nu}&=&\frac{1}{2}\,\,\Big(\varepsilon^{\mu}_{X}(+)
\bar{\varepsilon}^{\dagger\,\rho}_{\gamma}(-)
-\varepsilon^{\mu}_{X}(-)\bar{\varepsilon_{\gamma}}^{\dagger\,\rho}(+)
\Big)\,\,\varepsilon^{\dagger\,\nu}_{\Jpsi}(0)\,, \nonumber \\
\IP_{T}^{\mu \rho \nu}&=&\frac{1}{2}\,
\varepsilon^{\mu}_{X}(0)\,\,\Big(\bar{\varepsilon}^{\dagger\,\rho}_{\gamma}(+)
\varepsilon^{\dagger\,\nu}_{\Jpsi}(+)
-\bar{\varepsilon}^{\dagger\,\rho}_{\gamma}(-)\varepsilon^{\dagger\,\nu}_{\Jpsi}(-)\Big)\,,
\label{eq:projsum}
\ena     
where we use the Jacob-Wick convention for the helicity polarization 
four--vectors as written down in \cite{ab66}. The $z$--direction is defined 
by the momentum of the $\Jpsi$. The bars in the polarization four--vectors 
$\bar{\varepsilon}_{\gamma}^{\rho}(\lambda_{\gamma})$ of the photon are a 
reminder that the photon helicities are defined relative to the negative 
$z$--direction. In the present context it is important to take into account 
both parity configurations related by a helicity reflection in
the definition of Eq.~(\ref{eq:projsum}). In explicit form one has in the
X-rest frame

\bea
\varepsilon_{X\,\mu}(\pm) &=& \tfrac{1}{\sqrt{2}}\, \Big(0;\pm 1,i,0\Big),\qquad
p^\alpha = \Big(m_X; 0,0,0\Big),
\nn
\varepsilon_{X\,\mu}(0) &=& \Big(0;0,0,-1\Big), 
\nn[2ex]
\varepsilon^\dagger_{\Jpsi\,\nu}(\pm) &=& \tfrac{1}{\sqrt{2}}\, 
\Big(0;\pm 1,-i,0\Big),\qquad
q_1^\alpha = \Big(\tfrac{m_X^2+m_{\Jpsi}^2}{2m_X}; 0,0,|\vec q_2|\Big),
\nn
\varepsilon^\dagger_{\Jpsi\,\nu}(0) &=& 
\tfrac{1}{m_{\Jpsi}}\, \Big(|\vec q_2|;0,0,-\tfrac{m_X^2+m_{\Jpsi}^2}{2m_X}\Big)\,,
\nn[2ex]
\bar\varepsilon^\dagger_{\gamma\,\rho}(\pm) &=& 
\tfrac{1}{\sqrt{2}}\, \Big(0;\mp 1,-i,0\Big),\qquad
q_2^\alpha = |\vec q_2|\Big(1;0,0,-1\Big).
\label{eq:polarization}
\ena 
A convenient covariant representation of the projectors can be obtained in 
the form
\be 
\IP_{i}^{\mu \rho \nu}=h_{i}^{\mu' \rho' \nu'} {S_{X}^{(1)\mu}}_{\mu'}(p)\,
(-{g^{\rho}}_{\rho'})\,{S_{\Jpsi}^{(1)\nu}}_{\nu'}(q_1)\,,
\label{eq:proj}
\en
where
\bea
h_L^{\mu \rho \nu}&=&\frac{i}{2}\frac{m_{\Jpsi}}{(q_1\cdot q_2)^2}
\,\varepsilon^{\mu \rho q_1 q_2}\,q_2^\nu\,, \qquad 
h_T^{\mu \rho \nu} = -\frac{i}{2}\frac{m_X}{(q_1\cdot q_2)^2}\,q_2^{\mu}
\varepsilon^{\rho \nu q_1q_2}\, ,
\ena  
and where the massive propagator functions are given by $(V=X,\Jpsi)$ 
\be
{S_V^{(1)\,\,\alpha}}_{\alpha'}(p_V)=
-{g^\alpha}_{\alpha'}+ \frac{p_V^{\alpha}p_{V\,\alpha'}}{m_{V}^{2}}\,.
\en
The massive propagator functions are needed in the projectors 
Eq.~(\ref{eq:proj}) 
to project out the appropriate three--dimensional subspaces in the respective 
rest systems of the spin 1 particles. For the photon one exploits the gauge 
freedom to write the propagator function as $(-{g^{\rho}}_{\rho'})$. 
Note that the compact form (\ref{eq:proj}) is only obtained if one uses 
the summed form (\ref{eq:projsum}). 
The projection operators are orthonormal in the sense that 
$\IP_{i}^{\mu \rho \nu}\IP^{\dagger}_{j\,\mu \rho \nu}=-\frac{1}{2}\delta_{ij}$. 

The angular decay distribution in the decay 
$X(3872)\to \gamma + \Jpsi (\to \ell^{+}\ell^{-})$ is 
given by
\be
\frac{d\Gamma}{d\cos\vartheta}=BR(\Jpsi \to \ell^{+}\ell^{-})
\frac{1}{4\pi}\,\frac{1}{2S_X+1}\,
\frac{|\vec{q}_2|}{m_X^2}\,\Big( \frac{3}{4}\sin^2\vartheta \,\,|H_L|^2
+ \frac{3}{8}(1+\cos^2\vartheta)\, \,|H_{T}|^{2}\Big)\,,
\label{eq:angdist}
\en
where $\vartheta$ is the polar angle of either of the leptons $\ell^{\pm}$
relative to the original flight direction of the $\Jpsi$, all in the rest 
system of the $\Jpsi$. 

One can alternatively describe the transition in terms of the two
multipole amplitudes $A_{E1}$ and $A_{M2}$. The multipole amplitudes
are related to the helicity amplitudes via \cite{Cottingham:1976tw}
\be
A_{E1}=\frac{1}{\sqrt{2}}\Big(H_{L}-H_{T}\Big)\,, \qquad
A_{M2}=\frac{1}{\sqrt{2}}\Big(H_{L}+H_{T}\Big)\,.
\label{eq:mult}
\en
The corresponding projectors onto the multipole amplitudes are given by
\be
\IP_{E1}^{\mu \rho \nu}=\frac{1}{\sqrt{2}}\Big(\IP_{L}^{\mu \rho \nu}
-\IP_{T}^{\mu \rho \nu}\Big)\,, \qquad
\IP_{M2}^{\mu \rho \nu}=\frac{1}{\sqrt{2}}\Big(\IP_{L}^{\mu \rho \nu}
+\IP_{T}^{\mu \rho \nu}\Big)\,.
\label{eq:multproj}
\en

In Table 1 we have summarized the helicity and multipole amplitudes 
resulting from the relevant projections of the basic covariants
Eq.~(\ref{basis}). The entries can be seen to satisfy the
constraint equations Eqs.~(\ref{constraint1},\ref{constraint2}). 
The multipole amplitudes $A_{E1,M2}$ calculated from the gauge invariant 
structures
$K^{(i)}_{\mu \rho \nu}$ $(i=2,4,5,6)$ show the appropriate 
lowest-order power behavior $A_{E1}\sim |\vec{q}_{2}|$ and
$A_{M2}\sim |\vec{q}_{2}|^{2}$.

The leading $|\vec{q}_{2}|$ contribution to the angular decay distribution
proportional to $|A_{E1}|^{2}$ is thus given by 
$W(\cos\vartheta)\propto (3-\cos^{2}\vartheta)$. The next-to-leading
contribution proportional to $2{\cal R} (A_{E1}A^{*}_{M2})$ is down by one 
power of $|\vec{q}_{2}|$. The nonleading angular distribution is given by 
$W(\cos\vartheta)\propto (1-3\cos^{2}\vartheta)$\, (in the same units).
 
\begin{table}[ht]
        \begin{center}
\def\arraystretch{2}
        \begin{tabular}{|c|c||c|c|c|c|}
        \hline
\,\,\,$i$\,\,\, &
$K^{(i)}_{\mu\rho\nu}$ & $H_{L}^{(i)}
=\IP_{L}^{\mu\rho\nu} K^{(i)}_{\mu\rho\nu}$ &
$H_{T}^{(i)}=\IP_{T}^{\mu\rho\nu} K^{(i)}_{\mu\rho\nu}$ &
$A_{E1}^{(i)}=\IP_{E1}^{\mu\rho\nu} K^{(i)}_{\mu\rho\nu}$ &
$A_{M2}^{(i)}=\IP_{M2}^{\mu\rho\nu} K^{(i)}_{\mu\rho\nu}$ \\
\hline
 1 &
$\varepsilon_{\mu\rho\nu q_1}$ & $i m_{\Jpsi}$ & $-i\frac{m_X^2+m_{\Jpsi}^2}{2m_X}$ &
$\frac{i}{\sqrt{2}}\frac{(m_X+m_{\Jpsi})^2}{2m_X}$ &
$-\frac{i}{\sqrt{2}}\frac{2m_X}{(m_X+m_{\Jpsi})^2}|\vec q_2|^2$ \\
%\hline
 2 &
 $q_{1 \mu}\varepsilon_{\rho\nu q_1 q_2}$ & $0$ & $i m_X |\vec q_2|^2$ & 
$ - \frac{i}{\sqrt{2}} m_X |\vec q_2|^2$ &
$   \frac{i}{\sqrt{2}} m_X |\vec q_2|^2$ \\
%\hline
 3 &
$q_{1 \rho}\varepsilon_{\mu\nu q_1 q_2}$ & $0$ &$0$ & $0$ &$0$ \\
%\hline
 4 &
 $q_{2 \nu}\varepsilon_{\mu\rho q_1 q_2}$ & $i \frac{m_X^2}{m_{\Jpsi}}|\vec q_2|^2$ &
0 & $ \frac{i}{\sqrt{2}}\frac{m_X^2}{m_{\Jpsi}}|\vec q_2|^2$ &
$ \frac{i}{\sqrt{2}}\frac{m_X^2}{m_{\Jpsi}}|\vec q_2|^2$ \\
%\hline
 5 &
$\varepsilon_{\mu\rho\nu q_2}$ &   $i \frac{m_X}{m_{\Jpsi}}|\vec q_2|$ &
 $-i |\vec q_2|$ &  
$ \frac{i}{\sqrt{2}}\frac{m_X+m_{\Jpsi}}{m_{\Jpsi}}|\vec q_2|$ &
$ \frac{i}{\sqrt{2}}\frac{2 m_X}{m_{\Jpsi}(m_X+m_{\Jpsi})}|\vec q_2|^2$ \\
%\hline
 6 &
 \,\,\, $K^{(3)}_{\mu\rho\nu}-(q_1q_2)\, K^{(1)}_{\mu\rho\nu}$ \,\,\, & 
\,\,\,  $-im_Xm_{\Jpsi}|\vec q_2|$ \,\,\,                    &
 \,\,\, $i\frac{m_X^2+m_{\Jpsi}^2}{2}|\vec q_2|$ \,\,\,      &
 \,\,\, $-\frac{i}{\sqrt{2}}\frac{(m_X+m_{\Jpsi})^2}{2}|\vec q _2|$ \,\,\, &
 \,\,\, $ \frac{i}{\sqrt{2}}\frac{2m_X^2}{(m_X+m_{\Jpsi})^2}|\vec q_2|^3$ \,\,\,  
\\[2ex]
\hline
      \end{tabular}
        \end{center}
\label{tab:projectors}
\caption{Helicity and multipole projections of the basic tensors 
$K^{(i)}_{\mu\rho\nu}$. The tensors $K^{(i)}_{\mu\rho\nu}$\, $(i=2,4,5,6)$
are gauge invariant. \newline
They satisfy $q_2^\rho K^{(i)}_{\mu\rho\nu}=0.$
}
\end{table}

\end{document}